\documentclass[twocolumn,preprintnumbers,amsmath,amssymb]{revtex4}
\usepackage{graphicx}
\usepackage{amsmath,amssymb,latexsym}

\begin{document}

\title{Scaling and efficiency determine the irreversible evolution of a market}
 
\author{F. Baldovin}
\email{baldovin@pd.infn.it}
\affiliation{
Dipartimento di Fisica and
Sezione INFN, Universit\`a di Padova,\\
\it Via Marzolo 8, I-35131 Padova, Italy
}

\author{A. L. Stella}
\email{stella@pd.infn.it}
\affiliation{
Dipartimento di Fisica and
Sezione INFN, Universit\`a di Padova,\\
\it Via Marzolo 8, I-35131 Padova, Italy
}

\date{\today}

\begin{abstract}
$\;$\\
{\bf Classification:}\\
Physical Sciences/Statistics, Social Sciences/Economic Sciences\\
\\
{\bf Corresponding Author:}\\
Attilio Stella,\\
Dipartimento di Fisica and
Sezione INFN, Universit\`a di Padova,\\
{\it Via Marzolo 8, I-35131 Padova, Italy}\\
Tel: +39 049 827 7172\\
Fax: +39 049 827 7102\\
E-mail: stella@pd.infn.it\\
\\
{\bf Words and character counts:}\\
Abstract: 130 words\\
Total Character Count for the manuscript: 24989\\
\\
{\bf Abbreviations:}\\
Probability density function (PDF)\\
Dow Jones Industrial (DJI)\\
\\
{\bf Abstract:}\\
In setting up a stochastic description of the time evolution of a financial
index, the challenge consists  in devising a model compatible with
all stylized facts emerging from the analysis of financial time series
and providing a reliable basis for simulating such series.
Based on constraints imposed by market efficiency
and on an inhomogeneous-time generalization of standard
simple scaling, we propose an analytical model which accounts
simultaneously for empirical results like the linear decorrelation
of successive returns, the power law dependence
on time of the volatility autocorrelation function, and the multiscaling
associated to this dependence. In addition, our approach gives
a justification and a quantitative assessment of the irreversible character 
of the index dynamics. This irreversibility enters as a key ingredient in 
a novel simulation strategy of index evolution which demonstrates the
predictive potential of the model. 
\end{abstract}

\maketitle

$\;$\\

\newpage

$\;$\\
\newpage

{\bf Introduction}\\ 
For over a century it has been recognized \cite{bachelier_1} that the unpredictable time
evolution of a financial index is inherently a stochastic process.
However, in spite of many efforts 
\cite{mandelbrot_1,fama_1,black_1,engle_1,bollerslev_1,vassilicos_1,mandelbrot_3,andersen_1,lux_1,lebaron_1}, 
a unified framework for 
simultaneously understanding empirical facts 
\cite{mantegna_1,mantegna_2,bouchaud_2,cont_1,cont_2,stanley_1,yamasaki_1,lux_2}, 
such as the
non-Gaussian form and multiscaling in time of the distribution of returns,
the linear decorrelation of successive returns, and volatility clustering,
has been elusive. 
This situation occurs in
many natural phenomena, when strong correlations determine various forms of
anomalous scaling \cite{kadanoff_1,kadanoff_2,sethna_1,bouchaud_3,lu_1,scholz_1,kiyono_1,lasinio_1}. 
Here, by employing novel mathematical tools at the basis of a
generalization of the central limit theorem to strongly correlated
variables \cite{baldovin_1}, we propose a model of index evolution and
a corresponding simulation strategy which account for all robust
features revealed by the empirical analysis.

\begin{figure}
\includegraphics[width=0.98\columnwidth]{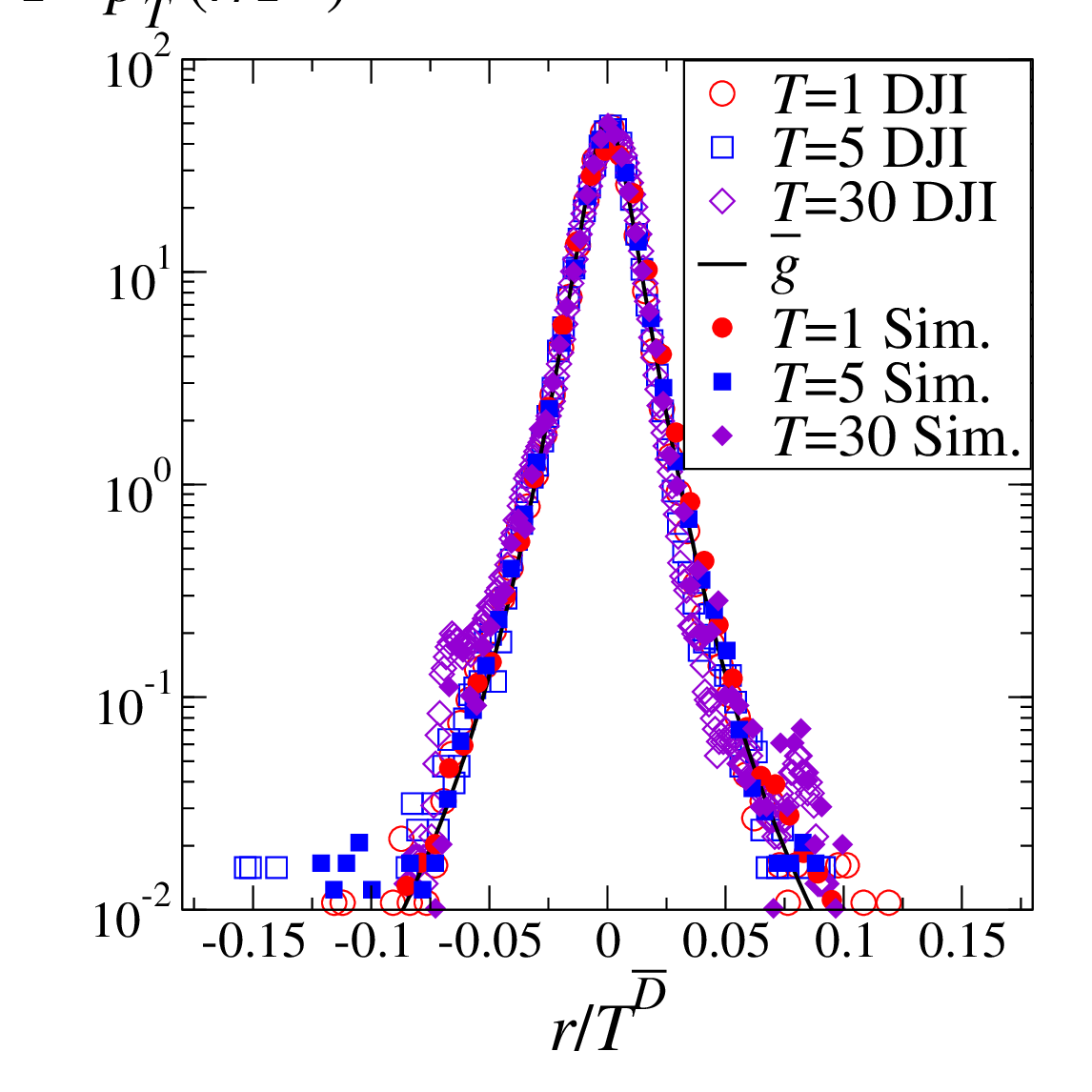}
\caption{
  Data collapse for $\overline{p}_T(r)$ ($T$ measured in days) sampled
  from a record of about $2.7\times10^4$ DJI daily closures (empty
  symbols). 
  The average daily trend 
  of the order of $10^{-4}$ has been subtracted. The collapse analysis
  furnishes the scaling function $\overline{g}$ reported as the full line and the
  scaling exponent $\overline{D}\simeq1/2$ (see also
  Fig. \ref{fig_mom} and Supporting Information).
  The full symbols report the data collapse for the results of a
  single simulation of the DJI history.  
}
\label{fig_scaling}
\end{figure}

Let $S(t)$ be the value of a given asset at time $t$.
The logarithmic return over the interval $[t,t+T]$ is defined as
$r(t,T)\equiv \ln S(t+T)-\ln S(t)$, where 
$t=0,1,\ldots$ and $T=1,2\ldots$, 
in some unit (e.g., day). 
From a sufficiently long 
historical series one can sample the empirical probability density function
(PDF) of $r$ over a time $T$, $\overline{p}_T(r)$, and the joint PDF of two 
successive returns $r_1 \equiv r(t,T)$ and $r_2 \equiv r(t+T, T)$, denoted by
$\overline{p}_{2T}^{(2)}(r_1,r_2)$. This joint PDF contains the
information on the correlation between $r_1$ and $r_2$ in the 
sampling. A well established
property 
\cite{mantegna_2,bouchaud_2,cont_1,cont_2}
is that, if $T$ is longer than tens of minutes,
the linear correlation vanishes:
$\int \overline{p}_{2T}^{(2)}(r_1,r_2) r_1 r_2 d r_1 d r_2
\equiv \langle r_1 r_2 \rangle_{\overline{p}_{2T}^{(2)}}=0$.
This is a consequence of the efficiency of
the market \cite{fama_1}, which quickly suppresses any arbitrage opportunity.
Another remarkable feature is that, within specific $T$-ranges, 
$\overline{p}_T$  
approximately assumes a simple scaling form
\begin{equation}
\overline{p}_T(r)= 
\frac{1}{T^{\overline{D}}}\;\;\overline{g}\left(\frac{r}{T^{\overline{D}}}\right),
\label{eq_scaling}
\end{equation}
where $\overline{g}$ and $\overline{D}$ are the scaling function and exponent,
respectively. 
Eq. (\ref{eq_scaling}) manifests self-similarity,
a symmetry often met in natural phenomena 
\cite{kadanoff_1,kadanoff_2,sethna_1,bouchaud_3,lasinio_1}: 
plots of $T^{\overline{D}}\overline{p}_T$ vs
$r/T^{\overline{D}}$ for different $T$'s collapse onto the
same curve representing $\overline{g}$.
We verify the scaling ansatz in Eq. (\ref{eq_scaling}) for the
Dow Jones Industrial (DJI) index using a dataset of more than
one century (1900-2005) of daily closures. This index is 
paradigmatic of market behavior and the considerable number of data
reduces sampling fluctuations substantially. In Fig. \ref{fig_scaling}
the collapse of the empty symbols is rather satisfactory 
(the explanation of the meaning of the full symbols in Fig. 1 is
given below).
The scaling function in Eq. (\ref{eq_scaling}) is non-Gaussian
\cite{mandelbrot_1,mantegna_1,mantegna_2,bouchaud_2,cont_1,cont_2,stanley_1}.
Although linear correlations vanish,
in the $T$-range considered $\overline{g}$ is
determined by the strong {\it nonlinear} correlations of the returns.
Only for $T>\tau_c$ (with $\tau_c$ of the order of the year)
successive index returns become  
independent and $\overline{p}_T$ turns Gaussian in force of the 
central limit theorem \cite{gnedenko_1}.

{\bf Results and discussion}\\
Our first goal is to establish  
up to what extent the assumption of simple scaling in
Eq. (\ref{eq_scaling}) 
does constrain the structure of the joint PDF
$\overline{p}_{2T}^{(2)}$.  
One must of course have
\begin{eqnarray}
&\int \overline{p}_{2T}^{(2)}(r_1,r_2) 
\delta (r-r_1-r_2) d r_1 d r_2=\overline{p}_{2T}(r),&\nonumber\\
&\int \overline{p}_{2T}^{(2)}(r_1,r_2) d r_2 =\overline{p}_T(r_1),&
\label{constraints}\\
&\int \overline{p}_{2T}^{(2)}(r_1,r_2) d r_1
=\overline{p}_T(r_2).&\nonumber
\end{eqnarray}
Indeed, the first line of Eqs. (\ref{constraints}) follows from
$r(t,2T)=r_1+r_2$.    
Furthermore, since the joint PDF $\overline{p}_{2T}^{(2)}$
is sampled from a sequence of time-translated
intervals of duration $2T$ along the historical series,
both the first and the second halves of all such intervals provide an adequate
sampling basis for $\overline{p}_T$.
This justifies the second and third lines of
Eqs. (\ref{constraints}). 
At this point we notice that the property $\langle r_1 r_2 
\rangle_{\overline{p}_{2T}^{(2)}} =0$ implies
that $\langle r^2 \rangle_{\overline{p}_{2T}}=
\langle (r_1 + r_2)^2 \rangle_{\overline{p}_{2T}^{(2)}}
=2\langle r^2 \rangle_{\overline{p}_T}$.
In force of \mbox{Eq. (\ref{eq_scaling})}, 
$\langle r^2 \rangle_{\overline{p}_T}
\sim T^{2\overline{D}}$. Hence, we obtain
$2 T^{2\overline{D}}=(2T)^{2\overline{D}}$, i.e. $\overline{D}=1/2$.
Remarkably, for all developed market indices,
$\langle r^2 \rangle_{\overline{p}_T}$ 
is found to scale consistently with a
$\overline{D}$ pretty close to $1/2$ \cite{di_matteo_1}.

\begin{figure}
\includegraphics[width=0.98\columnwidth]{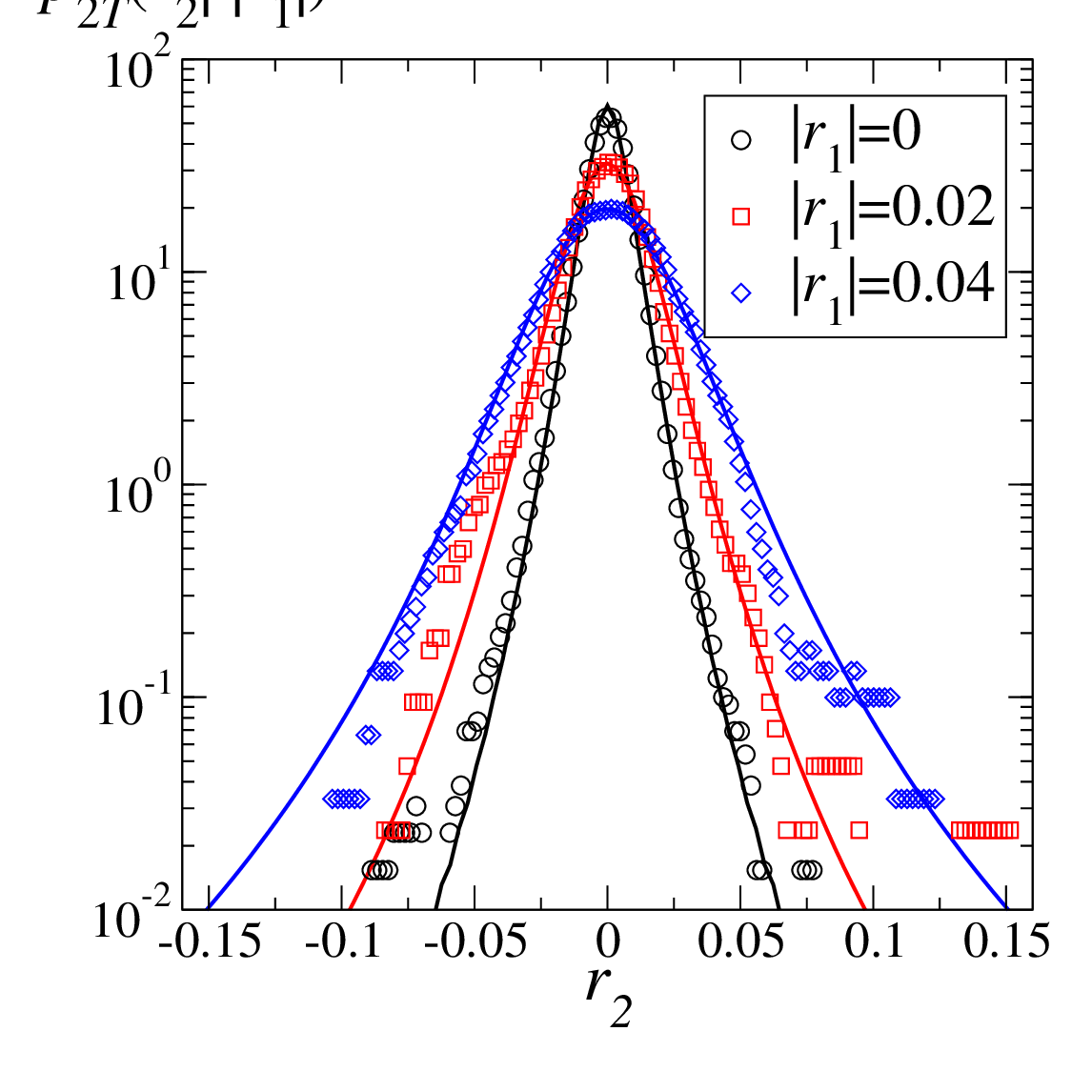}\\
\caption{
  Conditional probabilities of daily returns $r_2$ for different 
  values of $r_1$. 
  Empty symbols refer to the DJI data. 
  The continuous curves are the
  predictions of our theory for 
  $\overline{p}_{2T}^{(2)}(r_2||r_1|)\equiv
  [\overline{p}_{2T}^{(2)}(r_1,r_2)+
  \overline{p}_{2T}^{(2)}(-r_1,r_2)]/
  \int[\overline{p}_{2T}^{(2)}(r_1,r_2)+
  \overline{p}_{2T}^{(2)}(-r_1,r_2)]d r_2
  $.
  The absolute value of $r_1$ is introduced for reducing sample
  fluctuations.  
}
\label{fig_conditional}
\end{figure}

By switching to Fourier space in Eq. (\ref{constraints}), 
the notion of a novel, generalized product
operation allows to identify a solution for $\overline{p}_{2T}^{(2)}$ in terms of
$\overline{p}_T$ alone. 
While the ordinary multiplication of characteristic
functions (i.e., Fourier transforms) of Gaussian $\overline{p}_T$'s 
would yield trivially the correct $\overline{p}_{2T}^{(2)}$ in the case of 
independent successive returns \cite{gnedenko_1}, 
the generalized product is used here to 
take into account strong nonlinear correlations consistently with the anomalous 
scaling they determine (see Supporting Information) 
and can be seen to be at the basis of a novel central limit
theorem \cite{baldovin_1}. Our solution is strongly
supported by the remarkable consistency with the numerical results
and by the analogy with the independent case. In Fig. \ref{fig_conditional} we compare the
PDF of the return $r_2$ conditioned to a given absolute value of the return $r_1$,
as obtained through our solution (continuous lines), with the empirically sampled 
one (symbols). The agreement does not involve fitting parameters,
since $\overline{D}$ and 
those entering the assumed analytical form of 
$\overline{g}$ are already fixed 
in Fig. \ref{fig_scaling}.

\begin{figure}
\includegraphics[width=0.98\columnwidth]{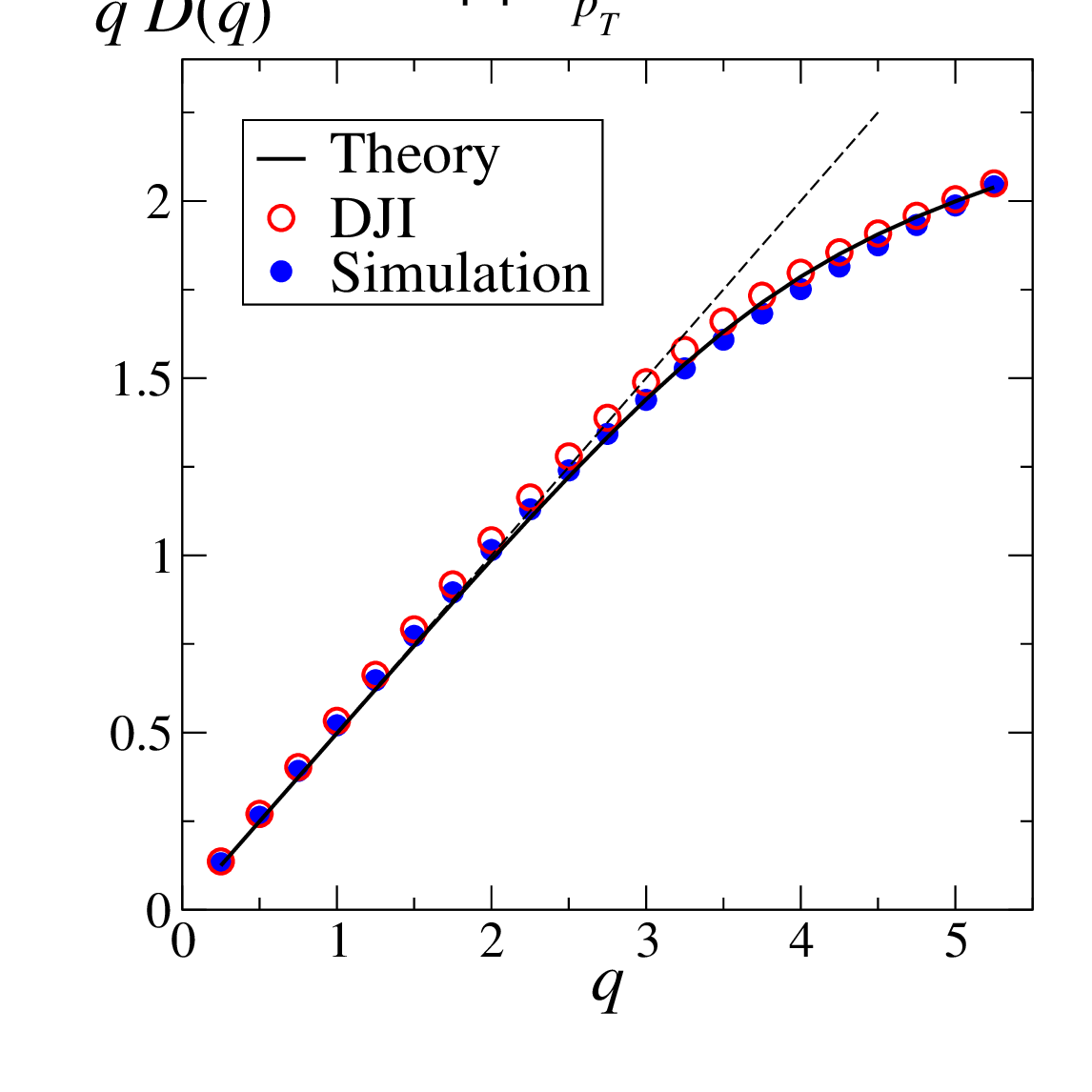}
\caption{
  Scaling exponent of the $q$-th moment of $\overline{p}_T$. 
  Empty (full) circles refer to the DJI data (simulation) of
  Fig. \ref{fig_scaling}.   
  The dashed line is $q/2$.
  Multiscaling is due to the deviation of $\overline{D}(q)$ from a
  constant value.
  The full line reports the time-averaged
  asymptotic ($\tau_c\gg1$) theoretical prediction based on
  Eq. (\ref{eq_average}).  
}
\label{fig_mom}
\end{figure}

At this point we must take into account that the simple scaling ansatz
in Eq. (\ref{eq_scaling}) is only approximately valid
\cite{vassilicos_1}.  
Indeed, a consequence of Eq. (\ref{eq_scaling}) is  
$\langle|r|^q\rangle_{\overline{p}_T}\sim T^{q\overline{D}}$, which we
exploited above for $q=2$.
However, a careful analysis reveals that the $q$-th moment exponent
deviates from the linear behavior $q\overline{D}\simeq q/2$ for 
$q\gtrsim3$ (empty circles in Fig. \ref{fig_mom}).
Like the linear behavior with slope $1/2$ observed for low-order
moments, 
this multiscaling effect is common to most indices \cite{di_matteo_1}.  
To explain this feature, we have to investigate the
relation between the empirical $\overline{p}_T$ and the stochastic
process generating the time series. 
If PDF's like $\overline{p}_T$ and $\overline{p}_{2T}^{(2)}$ were
directly describing such a process, this would be with stationary
increments. 
This assumption is legitimate 
only for sufficiently long times, larger than $\tau_c$.
Below, we identify in the interplay between scaling and
non-stationarity a precise mechanism accounting for the robust
features of $\overline{p}_T$ detected for $T\ll\tau_c$, including its
multiscaling.  

Let us indicate by
$p_{t,T}$ and $p_{t,2T}^{(2)}$ the ensemble PDF's corresponding to
$\overline{p}_{T}$ and $\overline{p}_{2T}^{(2)}$, respectively.
The additional dependence on $t$, the initial time
of the interval $[t,t+T]$, shows that we do not assume
stationarity for these PDF's.
We postulate that, within specific $T$-ranges (e.g., the one 
in Fig. \ref{fig_scaling}), 
$p_{0,T}$ obeys a simple scaling like that in
Eq. (\ref{eq_scaling}), but possibly with a $D$ and a
$g$ different from $\overline{D}$ and $\overline{g}$, respectively. 
One then realizes that this
scaling and the linear decorrelation of returns impose
on $p_{t,2T}^{(2)}$ 
constraints analogous to those for 
$\overline{p}_{2T}^{(2)}$ in Eq. (\ref{constraints}),
except for the third one,
which now reads 
\begin{equation}
\int p_{0,2T}^{(2)}(r_1,r_2) d r_1 \equiv p_{T,T}(r_2)
=p_{0,a T}(r_2).
\label{constraints_new}
\end{equation} 
This last condition tells us that, as a consequence of the nonlinear
correlations, the effective  time span of the marginal PDF obtained by
integrating $p_{0,2T}^{(2)}$ in $r_1$  must be renormalized by a factor $a$.
This factor is determined
again by consistency of the second moments scaling properties, as
above. 
Since now $\langle|r|^2\rangle_{p_{0,T}}\sim T^{2D}$, one gets from
Eq. (\ref{constraints_new})
\mbox{$a=(2^{2D}-1)^{1/2D}$}.
So, $D \neq 1/2$ implies $a \neq 1$ and thus
non-stationarity and irreversibility of the process. 
Similar functional relations 
hold for the PDF's of the magnetization of critical spin 
models upon doubling the system size and can be explained in that 
context by the renormalization group theory \cite{lasinio_1}.
Our generalized multiplication of characteristic functions allows us to
express $p_{t,2T}^{(2)}$ in terms of $p_{t,T}$ and to establish the
time-inhomogeneous scaling property 
\begin{equation}
p_{t,T}(r)=
\frac{1}{\sqrt{(t+T)^{2D}-t^{2D}}}\quad
g\left(\frac{r}{\sqrt{(t+T)^{2D}-t^{2D}}}\right).
\label{eq_scaling_new}
\end{equation}

It remains now to make explicit the link between the $p$'s and the
sampled $\overline{p}$'s and to determine $D$.
By construction, $\overline{p}_T$ is a $t$-average
of $p_{t,T}$. Since the time inhomogeneity of
$p_{t,T}$ must cross over into homogeneity for
$t$ exceeding $\tau_c$, we expect the following approximation 
\begin{equation}
\overline{p}_T(r) = \frac{1}{\tau_c} \sum_{t=0}^{\tau_c -1} p_{t,T}(r)
\label{eq_average}
\end{equation}
to hold.
Indeed, the history over which
$\overline{p}_T$ is sampled is much longer than $\tau_c$
and allows in principle also an indirect sampling of $p_{t,T}$ if we
simply assume $p_{t+\tau_c,T}(r)=p_{t,T}(r)$. 

In spite of the fact that Eq. (\ref{eq_scaling_new}) implies a simple
scaling exponent $D$ for $p_{0,T}$, Eq. (\ref{eq_average})
leads to the remarkable property that, independently of $D$, 
the low-$q$ moments of $\overline{p}_T$ approximately scale with
exponent $q/2$ as soon as $\tau_c\gg 1$. 
Moreover, if $D<1/2$, $\overline{p}_T$ displays a
multiscaling of the same type as that found empirically:
$\overline{D}(q)<1/2$ for the high-order moments. 
The matching of the theoretical predictions for the
multiscaling of $\overline{p}_T$ on the basis of
Eq. (\ref{eq_average}) with the empirical results is a first way of
identifying $D$.  
For the DJI, in Fig. \ref{fig_mom} we show that with
$D=0.24$ this matching is very satisfactory. 
The scaling functions of $p_{0,T}$ and $\overline{p}_T$ 
can also be shown to be simply related, once $D$ is 
known. 
We notice that the
observed  
multiscaling features of financial indices, which inspired 
multiplicative cascade models
\cite{mandelbrot_3,lux_2} in analogy with turbulence
\cite{kadanoff_1,kadanoff_2}, are explained here in terms of an
additive process possessing the time-inhomogeneous scaling
(\ref{eq_scaling_new}).    
\begin{figure}
\includegraphics[width=0.98\columnwidth]{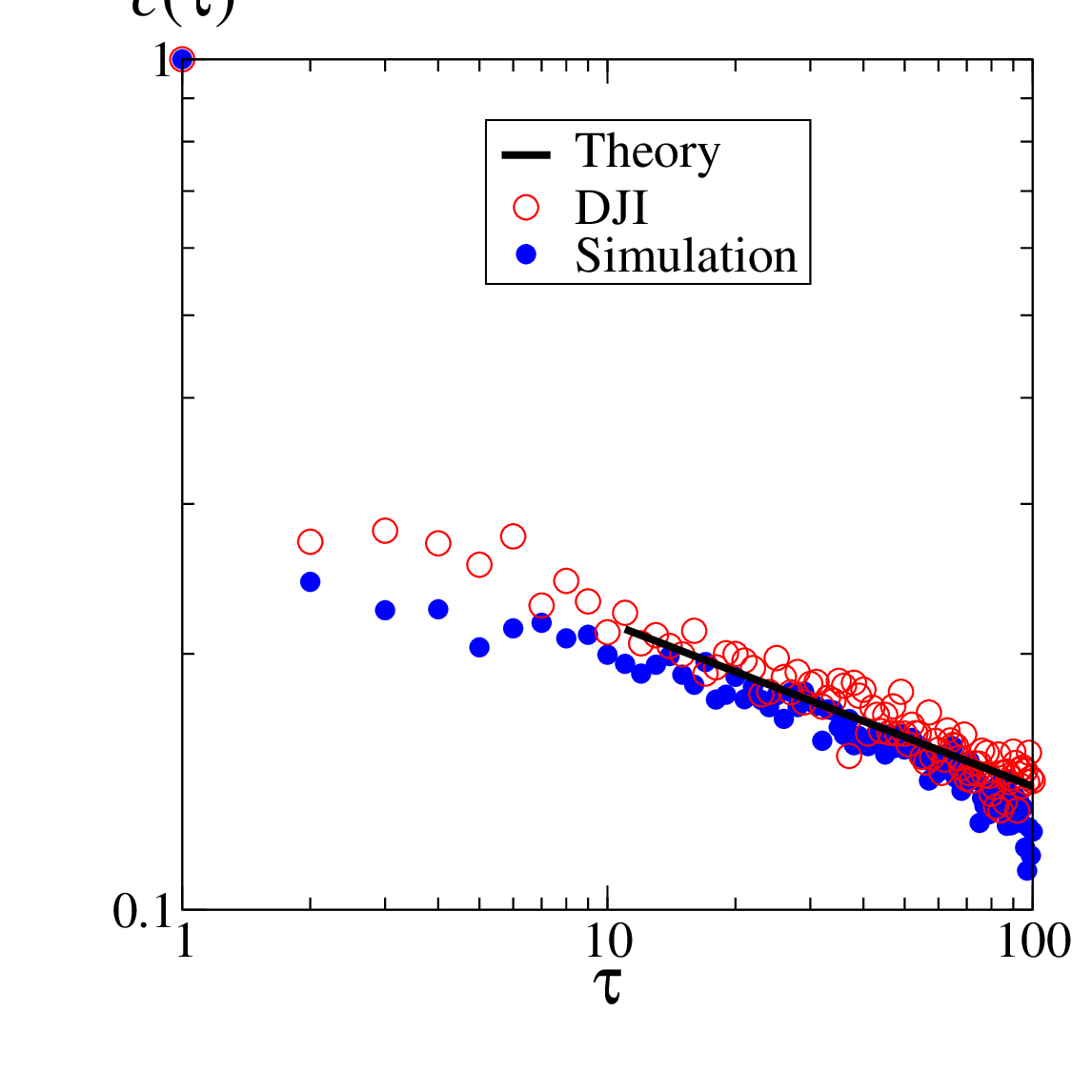}
\caption{
  Volatility autocorrelation at time separation $\tau$ (in days),
  \mbox{$
    c(\tau)\equiv\frac{\sum_{t=0}^{t_{max}} |r(t,1)| |r(t+\tau,1)| - 
      \sum_{t=0}^{t_{max}} |r(t,1)|\sum_{t=0}^{t_{max}} |r(t+\tau,1)|/t_{max}}
    {\sum_{t=0}^{t_{max}} |r(t,1)|^2 - [\sum_{t=0}^{t_{max}} |r(t,1)|]^2/t_{max}},
    \label{eq_correlation}
  $}
  where $t_{max}+\tau-1$ is the total length of the time series.
  Symbols are as in Fig. \ref{fig_mom}. 
  The simulation here is precisely the same we refer to in
  Figs. \ref{fig_scaling} and \ref{fig_mom}. 
  The full line gives the slope of the time-averaged asymptotic
  ($\tau_c\gg1$, $\tau\lesssim\tau_c$) model prediction for the volatility
  autocorrelation, superimposed to the data (see Supporting
  Information). 
}
\label{fig_correlation}
\end{figure}

The introduction of autoregressive schemes like ARCH \cite{engle_1}
marked an advance in econometrics and financial analysis
\cite{bollerslev_1,andersen_1}, and, more generally, in the
theory of stochastic processes. 
%as they introduced the possibility of simulating
%correlated processes whose long single realizations can be compared to
%the historical series. 
In an autoregressive simulation a number of parameters
weighting the influence of the past history on the PDF of the
following return must be fixed through some optimization procedure.   
By our approach, a generalization of $p_{t,2T}^{(2)}$ to the case of
$n$-consecutive intervals can be fully expressed just in terms of
$p_{t,T}$ and $D$. 
This is obtained by taking the inverse Fourier transform of our
solution for the characteristic function of the joint PDF (see
Supporting Information). In this way we can precisely calculate
the PDF which rules the extraction of the $i$-th return, $r_i$, 
giving as conditioning inputs the previous $m$ ones,
$r_{i-m},\ldots,r_{i-1}$. 
Consistently with our schematization in Eq. (\ref{eq_average}), the
existence of exogenous 
factors acting on the market can be taken into account by resetting
the width of the marginal PDF's with an (average) 
periodicity  equal to $\tau_c$ (see Supporting Information). 
The results for a single simulation with $m=100$, $\tau_c=500$, and
$D=0.24$ are illustrated by the full symbols in
Figs. \ref{fig_scaling} and \ref{fig_mom}. 
The coincidence of the scaling properties observed for the DJI with
those of our simulation furnish a second strong indication of the
validity of our approach and of the estimation of $D$.

The correctness of the value of $D$ can be further checked by
considering the volatility autocorrelation
function at time-separation $\tau$ (Fig. \ref{fig_correlation}).
A well established fact \cite{mantegna_1,mantegna_2,bouchaud_2,cont_1,cont_2,lux_2}
is its power law decay 
$c(\tau) \sim \tau^{-\beta}$ for $ \tau < \tau_c$,
with $\beta\simeq 0.2$ for the DJI.
This behavior is not reproduced by routine simulation
methods in quantitative finance like GARCH \cite{bollerslev_1}
and requires the introduction of more sophisticated, fractional integration
techniques \cite{andersen_1}.
The full characterization of the joint PDF of $n$-consecutive
returns allows us  to obtain a model expression for $c(\tau)$, which again
takes into account the non-stationarity of the process
(see Supporting Information). 
Such an expression
behaves asymptotically as a power of $\tau$ 
with an exponent depending on
$D$ ($c(\tau)$ is constant for $D=1/2$ and decays for $D<1/2$). 
In particular, with $D=0.24$ both the model asymptotic expression and
the results of our simulation procedure furnish a nice
agreement with the exponent $\beta\simeq0.2$ observed for the 
DJI index (Fig. \ref{fig_correlation}).
Thus, the algebraic volatility autocorrelation function decay 
is reproduced by our scheme and provides a second
criterion to fix consistently the anomalous scaling exponent $D$.

Our approach is based on two postulates: inhomogeneous-time scaling
and the vanishing of linear return correlations. 
These symmetries lead, in an unambiguous, deductive manner, to a
model for the underlying 
stochastic process determining market evolution.
Of course, the results follow only when
the postulates are valid
and we have shown that within specific time-ranges the consequences of 
these postulates are in remarkable agreement with the data. 
Major advances in
understanding critical phenomena worked in a similar vein
decades ago \cite{kadanoff_1}, when the scaling assumptions allowed to
establish links between seemingly disparate phenomena and put the
basis for the development of renormalization group theory
\cite{lasinio_1}.  
So far, the coexistence of anomalous scaling with
the requirement of absence of linear correlation imposed by economic
principles has been regarded as an outstanding open problem in the
theory of stochastic processes. 
We believe that our solution could be relevant for developments in this
field, as well as for describing scaling behaviors of other complex
systems \cite{kadanoff_1,kadanoff_2,sethna_1,bouchaud_3,lu_1,scholz_1,kiyono_1}. 

{\bf Acknowledgments}\\
We thank J.R. Banavar for useful suggestions and encouragement.

\end{document}